# Electronic transport in amorphous Ge$_2$Sb$_2$Te$_5$ phase-change memory line cells and its response to photoexcitation

A. Talukder, M. Kashem, M. Hafiz, R. Khan, F. Dirisaglik, H. Silva, and A. Gokirmak
*Department of Electrical and Computer Engineering, University of Connecticut, Storrs, Connecticut 06269, USA*

We electrically characterized melt-quenched amorphized Ge$_2$Sb$_2$Te$_5$ (GST) phase-change memory cells of 20 nm thickness, ~66 - 124 nm width and ~100 - 600 nm length with and without photoexcitation in 80 - 275 K temperature range. The cells show distinctly different current-voltage characteristics in the low-field (< ~19 MV/m), with a clear response to optical excitation by red light ($\lambda_c \approx 613$ nm), and high-field (> ~19 MV/m) regimes, with very weak response to optical excitation. The reduction in carrier activation energy with photoexcitation in the low-field regime increases from ~10 meV at 80 K to ~50 meV at 150 K (highest sensitivity) and decreases again to 5 meV at 275 K. The heterojunctions at the amorphous-crystalline GST interfaces at the two sides of the amorphous region lead to formation of a potential well for holes and a potential barrier for electrons with activation energies in the order of 0.7 eV at room temperature. The alignment of the steady state energy bands suggests formation of tunnel junctions at the interfaces for electrons and an overall electronic conduction by electrons. When photoexcited, the photo-generated holes are expected to be stored in the amorphous region, leading to positive charging of the amorphous region, reducing the barrier for electrons at the junctions and hence the device resistance in the low-field regime. Holes accumulated in the amorphous region are drained under high electric fields, hence the cells' response to photoexcitation is significantly reduced. These results support electronic origin of resistance drift in amorphous GST.

Phase-change memory (PCM), also known as PCRAM (phase-change random access memory), OUM (ovonic unified memory), and C-RAM (chalcogenide RAM), stands at the forefront of contemporary memory technologies, offering a compelling solution for high-density non-volatile memory operation with its 2-terminal resistive technology[1–3]. PCM exploits the unique behavior of chalcogenide glasses, Ge$_2$Sb$_2$Te$_5$ (GST) being the most commonly used material, with ability to switch rapidly and repeatedly between highly conductive crystalline and highly resistive amorphous phases through the application of suitable electrical pulses[1–7]. The remarkable endurance of up to ~$10^{12}$ cycles[6,8,9] with potential improvement projected to $10^{15}$ cycles[9] and the demonstration of large resistivity contrast[10,11] highlight its promises in the memory market.

PCM cells experience very high temperatures (~1000 K), thermal gradients (~10 K/nm), current densities ($10^7$ A/cm$^2$), and electric fields (~30 MV/m) within small dimensions (down to 5 nm scale) over short timescales (10 ns – 1 μs)[12,13] during reset and set operations[14]. Even though the cell operation is rather complicated compared to conventional electronic devices and the behavior of materials is not fully understood yet, PCM is demonstrated to be a reliable and cost-effective non-volatile memory technology that can be integrated with CMOS at the back-end-of-the-line[15].

One important question that remains regarding PCM is on the physical phenomena that give rise to an upward resistance drift in amorphous phase-change materials, approximately following a power law behavior[16–19]:

$$R = R_0 \left(\frac{t}{t_0}\right)^\nu \quad (1)$$

where $R$ and $R_0$ are the cell resistances at time $t$ and $t_0$, respectively and $\nu$ is the drift coefficient. Although resistance drift in GST is generally attributed to thermally activated structural relaxation of the material[20–34], low temperature resistance drift (Fig. 1b) and slow photo-response observations[35] and theoretical predictions[36,37] suggest an electronic origin instead, based on charge relaxation[36,37].

In this work, we performed current-voltage (I-V) measurements on amorphized (reset) PCM line cells with and without photoexcitation in 80 - 275 K temperature range after stabilization with high electric field stresses, which substantially accelerates resistance drift and brings it to a stop[39–41], and construct the band-diagrams to analyze charge trapping and electronic transport to explain the observed behavior.

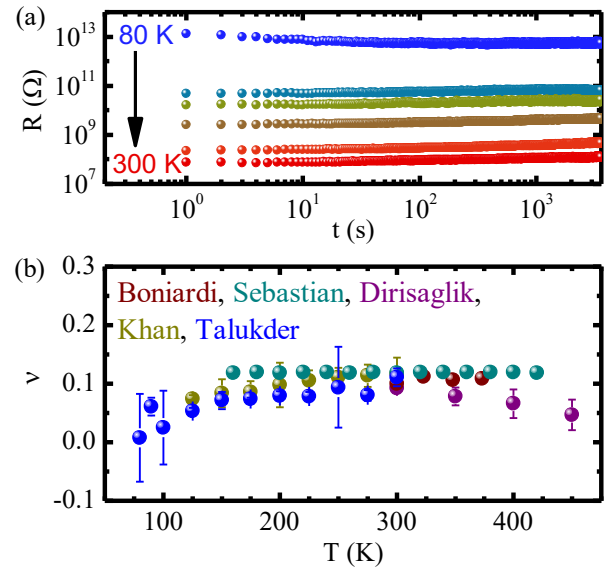

FIG. 1. (a) Amorphous resistance drift measurements performed within ~1 s of amorphization (shown for select temperatures in 80 – 300 K). (b) Reported drift coefficients from our group and others[28,29,35,38]. Our results are from 38 cells with W × L ≈ 66 – 88 nm × 220 – 388 nm and are extracted from linear fits of log($R/R_0$) vs log($t/t_0$) for the first 600 points after amorphization.

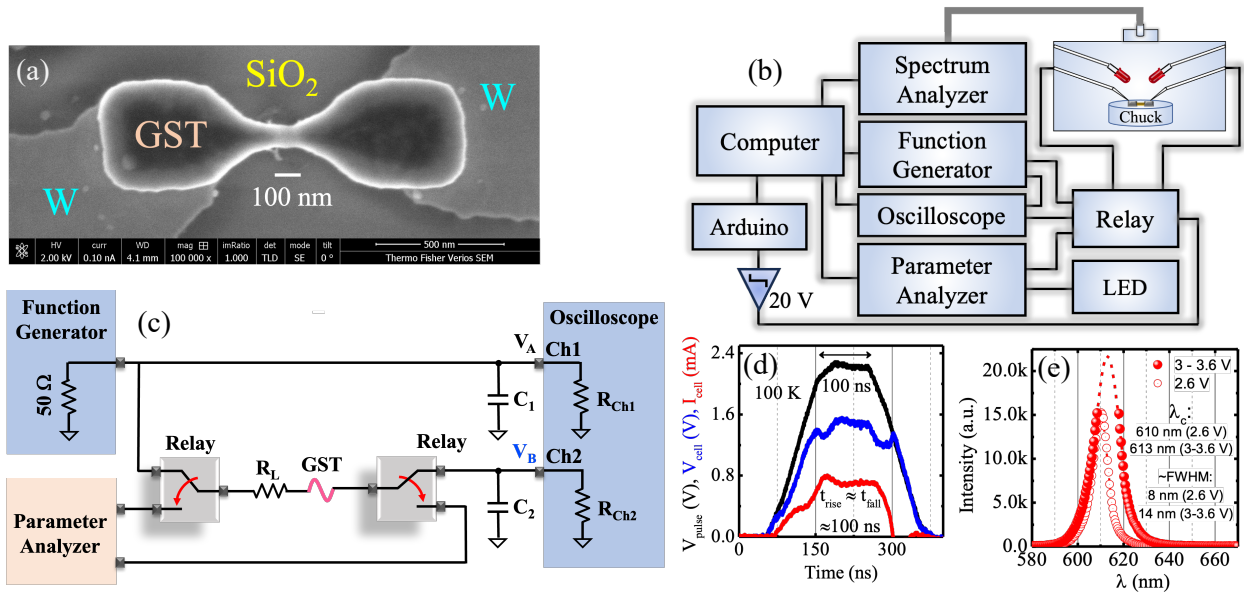

FIG. 2. (a) Scanning electron microscopy (SEM) image of a GST line cell with Tungsten bottom contacts. (b)-(c) The electrical measurement setup and its equivalent circuit diagram consisting of a 2-way relay between a function generator, an oscilloscope, and a parameter analyzer. (d) ~100 ns amorphization pulse and the resulting cell current for a cell with W × L × t ≈ 66 nm × 260 nm × 20 nm at 100 K. (e) Spectrum of a red light emitting diode (LED) under 2.6 and 3 - 3.6 V biases at 100 K with corresponding Gaussian fits.

The two terminal line cells (Fig. 2a) used for the experiments have 250 nm thick tungsten bottom metal contacts on 600 nm thermally grown $SiO_2$. A 20 nm layer of GST was deposited by co-sputtering from elemental targets at room temperature resulting in as-deposited amorphous phase, capped by a 15 nm layer of $SiO_2$ deposited by plasma enhanced chemical vapor deposition (PECVD) to prevent any oxidation and/or evaporation during operation. The line cells are patterned using photolithography and reactive ion etching (RIE). Details of the device fabrication processes are described by Dirisaglik, et al[42,43]. The cells are annealed at 675 K for ~20 minutes to crystallize GST to the stable hexagonal close pack (hcp) phase. The electrical experiments are conducted using a cryogenic probe station (Janis ST-500-UHT) integrated with a computer controlled (LabVIEW) setup utilizing a 2-channel relay (Agilent 16440A), a parameter analyzer (Agilent 4156C), an arbitrary waveform generator (Tektronix AFG 3102), a digital oscilloscope (Tektronix DPO 4104) and an Arduino board (Fig. 2b,c). The parameter analyzer is used for I-V characterization and to monitor the resistance drift after amorphization, while the arbitrary waveform generator and the oscilloscope are used to apply the amorphization pulses and monitor the cell voltage and current during the pulse (Fig. 2d). The 2-channel relay is controlled through the Arduino board and enables fast switching (< 1 s) and low-leakage (< 10 fA) measurements compared to a delay of ~100 s in our previous measurements[35].

The amorphized GST line cells are electrically stressed and stabilized using dc high-field sweeps[39,40] after low-voltage drift measurements. The cells are then characterized by I-V measurements at 80, 90, and 100 – 300 K in 25 K intervals in dark, and then under illumination from a red ($\lambda_c \approx 613$ nm) LED installed inside the cryogenic chamber (Fig. 2b,e). The current compliance is kept at ~50 nA at T ≤ 250 K and ~100 nA at T = 275 - 300 K, hence no significant self-heating of the cells is expected[40].

I-V characteristics of the amorphous GST cells exhibit two distinct exponential behaviors at low field (< ~19 MV/m) and high field (< ~19 MV/m) regimes at all temperatures[39,40] regardless of photoexcitation (Fig. 3a-c). Sensitivity to photoexcitation, S = ($I_{dark}$ - $I_{light}$)/$I_{dark}$, (Fig. 3d) is very strong in the low-field regime at lower temperatures and weakens at higher temperatures and in the high-field regime.

Amorphous GST is a lightly p-type semiconductor, established by thermoelectric measurements[44], with a room temperature bandgap of ~1 eV[45], that is expected to exhibit hopping transport[46–52]. However, electronic transport in small-scale devices is significantly impacted by their interfaces with the contact material. The contacts can be Ohmic or rectifying depending on the bandgap, electron affinity and the Fermi level of the materials at the interface. Here we construct the expected energy-band diagram for a 300 nm amorphized section (a-GST) in between two 200 nm crystalline GST sections (c-GST) (Fig. 4) using parameters we obtained from the literature: electron affinity of c-GST and a-GST as 4.18 eV and 4.99 eV[53]; bandgap ($E_g$) of c-GST and a-GST as 0.35 eV[54,55] and 1 eV[56]. By assuming a carrier concentration of $1\times10^{20}$ cm$^{-3}$ and $1\times10^{17}$ cm$^{-3}$ for c-GST and a-GST[52,57,58], respectively, the Fermi level ($E_F$) is calculated to be ~0.035 V below the valence band edge ($E_V$) for degenerate p-type c-GST and ~0.15 eV above $E_V$ for slightly p-type a-GST (Fig. 4a). The equilibrium condition is expected to be established after completion of charge exchanges (Fig. 4b), where Fermi levels are aligned. The resulting heterostructure shows substantial barriers for holes at the left and right interfaces and tunnel barriers for electrons due to substantial difference in the electron affinity values. The fixed negative charges at the two ends of the amorphous region are due to emission of holes from the hole-traps which are lost to the crystalline contact regions. As the trapped holes continue to get de-trapped over time, their time-to-escape the potential well gets longer as the well gets deeper. The gradual distortion of the conduction-band edge increases the activation energy for the electrons at the two contacts, reducing their rate for thermionic emission over the barrier, leading to increased resistance. This increase in resistance, which slows down in time, is in line with the power-law behavior observed for resistance drift[16–19].

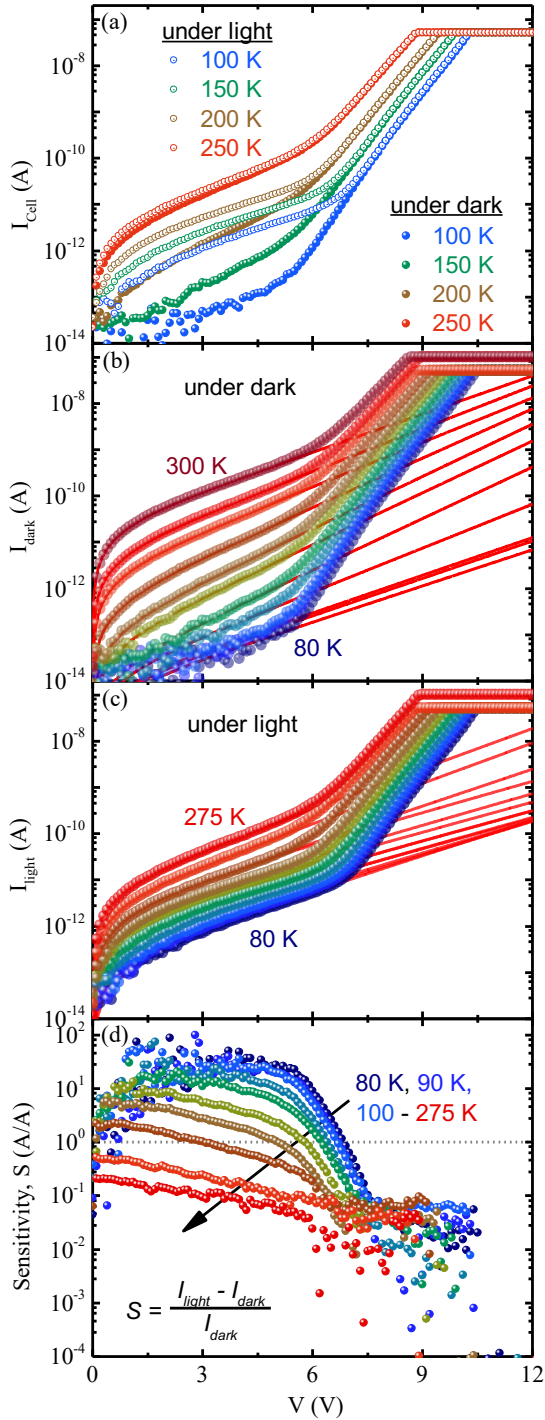

FIG. 3. (a) Responses under dark and light for select temperatures for a GST cell of dimension W × L ≈ 66 × 260 nm. (b) I-V curves with fits at T = 80 – 300 K under dark and (c) at 80 - 275 K under light. I-V sweeps are performed using 0 to 15 V and continuously from 15 to 0 V (shown up to 12 V) sweeps with ~50 - 100 nA current compliance under dark and subsequently under light. The fits made in the low-field regime are based on the hopping transport model and are extrapolated to the full range. (d) Sensitivity to photoexcitation calculated from the difference in responses under light and under dark shows declining photo-response in the high field regime for V > ~5.4 V. Sensitivity also goes down as the temperature increases and becomes insignificant at T > ~225 K.

Under a mild bias (0.1 V for the example case in Fig. 4c), the symmetry between forward transmission and reverse transmission is broken. The barrier height for forward and reverse transmission is modulated by the applied bias and hopping transport is expected to take place leading to:

$$I_{low\text{-}field} = \underbrace{I_{0,const}\, e^{-\frac{E_a}{kT}}}_{I_0} \left(e^{\alpha_1 V} - e^{-\alpha_2 V}\right) \quad (2)$$

where $I_{0,const}$ is the low-field pre-factor (a constant depending on the device geometry), $E_a$ is the zero-bias activation energy assuming an Arrhenius behavior for thermionic emission from the trap-sites and $\alpha_1$ and $\alpha_2$ are the coefficients defining the modulation in the potential barrier with applied bias for forward and reverse transmissions.

Under illumination with sufficiently large photon energy, electron-hole pairs will be the generated, which are then separated by the built-in field, electrons getting collected by the two c-GST contacts and holes accumulating in the amorphous region. The positive charging of the amorphous region leads to lowering of the potential barrier for both electrons (increasing conductivity) and holes (decreasing time-to-escape). With application of a high external electric-field (5.4 V for the example case in Fig. 4d), the potential barrier becomes significantly modified, leading to high-level injection of electrons and drainage of the holes, which are normally expected to be stored in the amorphous region for a long time. This, which also accelerates de-trapping of holes from the hole traps, leads to a substantial increase in the current flow regardless of photoexcitation (there is no longer a potential well to store the photo-generated holes) and accelerates resistance drift. As the temperature increases, the kinetic energy of the carriers increases and the bandgap decreases[59,60] leading to reduced time-to-escape for trapped holes (increased resistance drift and reduced response to photo-generated holes), and increased carrier injection rate in the low-field regime (Fig. 3c).

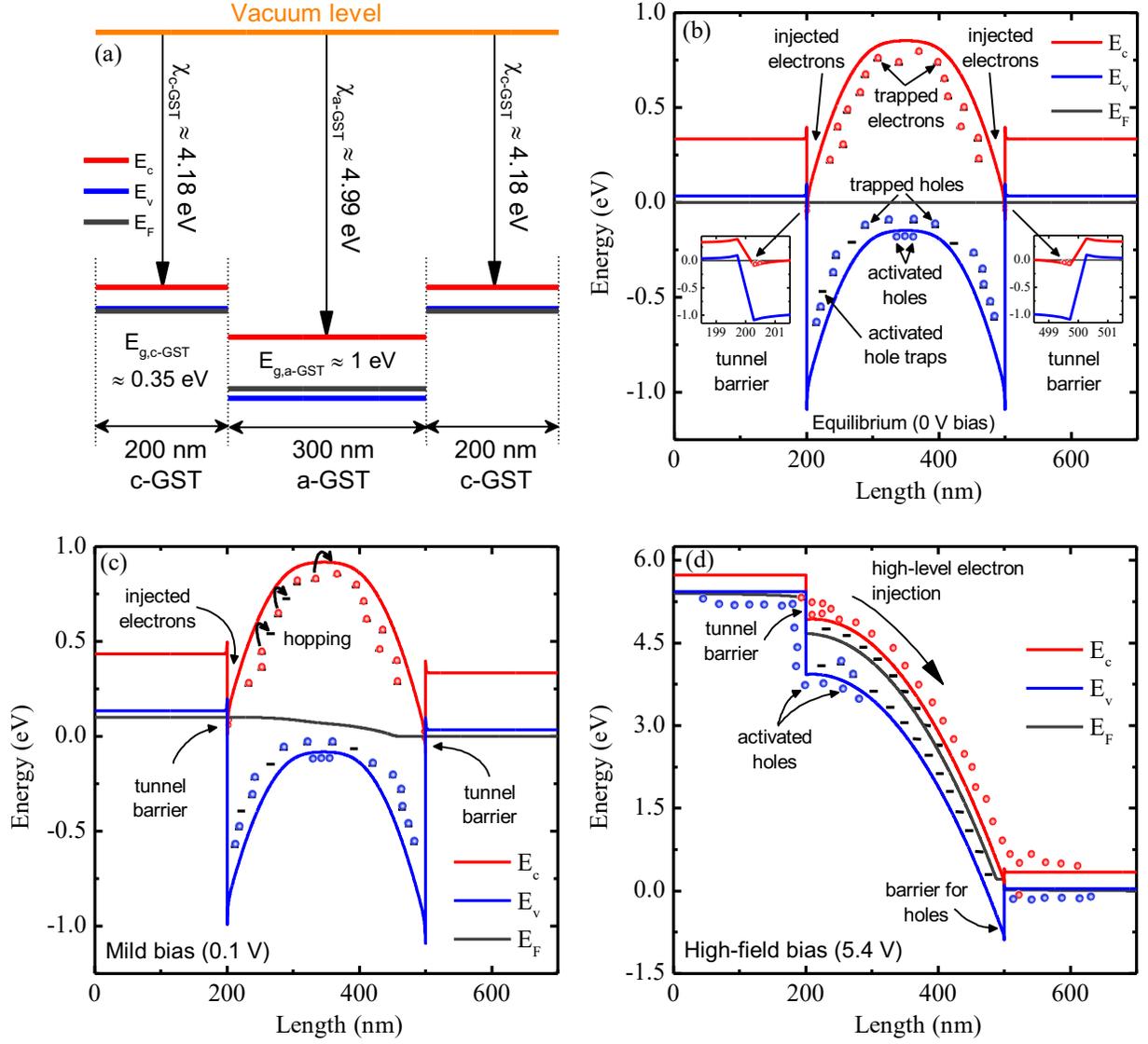

FIG. 4. Approximate energy-band diagrams of an amorphized GST wire with crystalline GST on both sides of the amorphized region (a) in flat band condition before forming the junctions, (b) in equilibrium case under zero bias, (c) under mild bias of 0.1 V, and (d) under a strong bias of 5.4 V. The a-GST/c-GST interfaces form substantial barriers for holes and tunnel barriers for electrons. Holes trapped in the central potential well are expected to escape under strong electric field and the electrons injected into the conduction band are expected to give rise to significant current flow.

The ratio of $I_0$ with ($I_{0,light}$) and without ($I_{0,dark}$) photoexcitation yields the difference in the activation energies between the two cases (Fig. 5):

$$\Delta E = E_{a,dark} - E_{a,light} = kT\, ln(I_{0,light}/I_{0,dark}) \quad (3)$$

We observe the change in the activation energy to increase from ~10 meV at 80 K to ~50 meV at 150 K and decrease to ~5 meV at 275 K. These relatively small perturbations are expected to be the impact of the change in the overall potential profile on the barriers between the rate-limiting hopping sites, where the trap-to-trap distances are expected to be in the order of 5 nm[40,61].

The peak in sensitivity to photoexcitation at ~150 K (Fig. 5b) may be related to possible match between the energy of the red photons (~2 eV) and the bandgap of a-GST at ~150 K. Absorption at lower temperatures is expected to be lower due to the increase of the a-GST bandgap beyond 2 eV. At higher temperatures, as bandgap of a-GST declines to ~1 eV at 300 K, the excess kinetic energy in the photo-generated holes may allow the holes escape from the potential well (Fig. 4c).

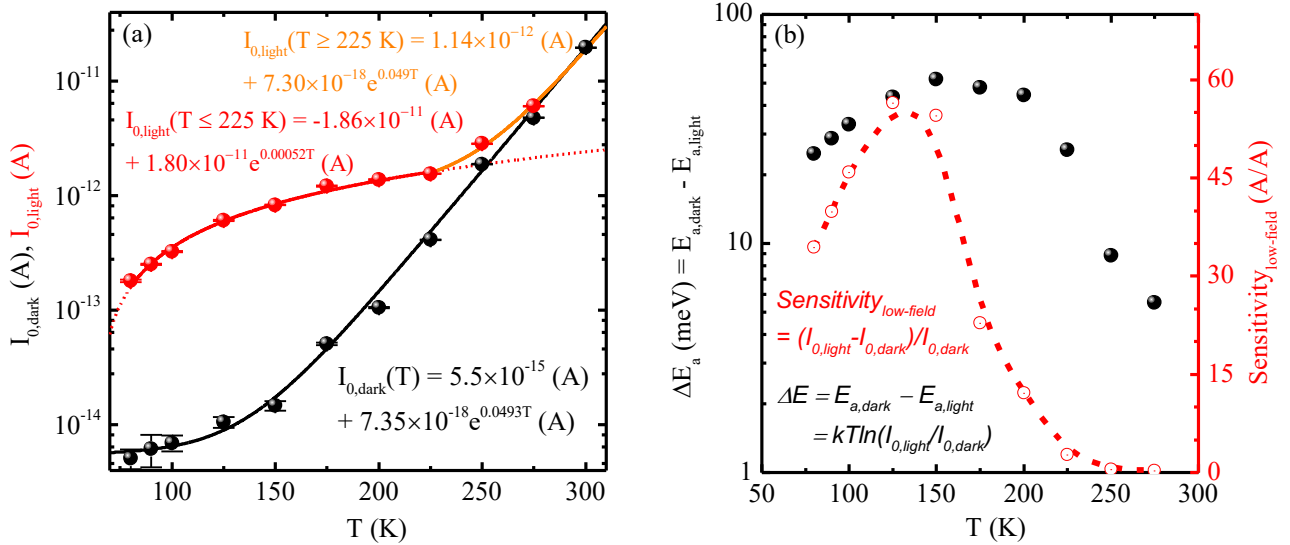

FIG. 5. (a) Temperature dependence of low-field current pre-factors under dark ($I_{0,dark}$) and under light ($I_{0,light}$) with exponential fits. The pre-factor under light is fitted with piecewise exponentials, one up to 225 K and then from 225 K to 275 K. (b) Difference in the activation energies under dark and under light increases with increasing temperature and starts coming down at ~150 K with further increase in temperature. Sensitivity calculated from the low-field current pre-factors show a decreasing trend at ~150 K suggesting a possible relation of incident light energy to the bandgap of the amorphous GST material.

In summary, in our experiments on amorphized $Ge_2Sb_2Te_5$ phase-change memory line cells, we observe two distinct characteristics at low electric field and high electric field regimes. The cells show high sensitivity to photoexcitation in the low-field regime, which peaks around 150 K for photoexcitation with red light (~2 eV). The energy-band diagrams we constructed using the published material parameters show a device level potential profile that limits high-level injection of electrons in the low-field regime and a potential well for holes that can store activated holes for long durations. Based on this analysis we conclude that electrons, which are minority carriers in a-GST, are the dominant carriers for current conduction between the two crystalline GST regions. Electrons are injected to the amorphous side of the amorphous-crystalline interfaces from the valance-band of the crystalline regions through tunnel barriers. The potential barrier for the electrons in a-GST forms over a relatively long duration after melt-quench, with the process slowing down from the first tens of nanoseconds[18] to many months[62] with the deepening of the potential well for holes in time, observed as resistance drift. The constructed energy-band diagram also explains the very slow response of the a-GST cells to photoexcitation, substantially increased current in the high-field regime, acceleration of resistance drift with application of high-field electric stresses. Hence, resistance drift can be explained by a purely electronic process and can be mitigated through device and waveform engineering. Mitigation of resistance drift in phase-change memory will enable multi-bit-per-cell storage for substantial reduction in cost-per-bit for PCM as well as reliable implementations of PCM as hardware accelerators in artificial intelligence (AI)-based applications.

A. Talukder and H. Silva worked on the electrical characterization, data analysis and manuscript preparation, supported by the US National Science Foundation (NSF) through award #1710468. M. Kashem, M. Hafiz, R. Khan and A. Gokirmak worked on modeling, design of experiments, data analysis and manuscript preparation supported by NSF grant #1711626. The devices were fabricated by F. Dirisaglik at IBM Watson Research Center under a Joint Study Agreement.